\documentclass[prd,twocolumn,nofootinbib,showpacs,superscriptaddress]{revtex4}

\usepackage{amsfonts}
\usepackage{amsmath}
\usepackage{amssymb}
\usepackage{amsthm}
\usepackage{bm}
\usepackage{dcolumn}
\usepackage{epsfig}
\usepackage{graphicx}
\usepackage{graphics}
\usepackage[latin1]{inputenc}
\usepackage{latexsym}
\usepackage{rotating}
\usepackage{hyperref}

\newcommand{\beqs}{\begin{equation*}}
\newcommand{\beq}{\begin{equation}}

\newcommand{\eeqs}{\end{equation*}}
\newcommand{\eeq}{\end{equation}}

\newcommand{\beqas}{\begin{eqnarray*}}
\newcommand{\beqa}{\begin{eqnarray}}

\newcommand{\eeqas}{\end{eqnarray*}}
\newcommand{\eeqa}{\end{eqnarray}}

\newcommand{\eq}[2]{\begin{equation} #1 \label{#2} \end{equation}}

\newcommand{\eps}{\varepsilon}
\newcommand{\al}{\alpha}

\newcommand{\ga}{\gamma}
\newcommand{\de}{\delta}

\newcommand{\Ga}{\Gamma}
\newcommand{\De}{\Delta}

\newcommand{\blist}{\begin{itemize}}

\newcommand{\elist}{\end{itemize}}

\providecommand{\href}[2]{#2}

\DeclareFontFamily{OT1}{rsfs}{}
\DeclareFontShape{OT1}{rsfs}{m}{n}{ <-7> rsfs5 <7-10> rsfs7 <10->rsfs10}{} 
\DeclareMathAlphabet{\mycal}{OT1}{rsfs}{m}{n}

\begin{document}
\title{Exact relativistic viscous fluid solutions in near horizon extremal Kerr background}

\author{Daniel Grumiller}
\affiliation{Institute for Theoretical Physics, Vienna University of Technology, Wiedner Hauptstr. 8-10/136, A-1040 Vienna, Austria, Europe}

\author{Ana-Maria Piso}
\affiliation{Kavli Institute for Astrophysics, Massachusetts Institute of Technology, 77 Massachusetts Ave., Cambridge, MA 02139, USA}

\date{\today}

\preprint{TUW-09-12}

\begin{abstract}

Realistic accretion disk models require a number of ingredients, including viscous fluids, electromagnetic fields and general relativistic corrections. Close to the innermost stable circular orbit (ISCO) the latter can be appreciable and (quasi-)Newtonian approximations become unreliable. This is particularly true for nearly extremal black holes like GRS 1915+105, where the ISCO almost coincides with the black hole horizon. To describe the physics close to the ISCO adequately in a simplified model we approximate the nearly extremal Kerr geometry by the near-horizon extremal Kerr geometry and construct in this background relativistic viscous fluid solutions with electromagnetic fields. We discuss some applications of our solutions and possible relations to the Kerr/CFT correspondence.

\end{abstract}


\maketitle


\noindent{\small {\em Our understanding of black hole accretion rests on analytic models.}\qquad\hfill Marek Abramowicz \cite{Abramowicz:2008bk}}

\section{Introduction}
\label{sec:1}

John Wheeler coined the term ``Black Hole'' more than four decades ago \cite{Wheeler:1967}. At that time Black Holes (BHs) were considered as rather esoteric objects of purely theoretical interest and little physical relevance (for a textbook on BHs and a history see Ref.~\cite{Frolov:1998}). The current assessment of the role of BHs in physics has changed dramatically, for several independent reasons. BHs are now used in various areas of physics, far beyond the original realms of application envisaged in the pioneering first decade of BH research, including quantum chromodynamics and condensed matter physics. We focus here on BHs in astrophysics and, to a lesser extent, on BHs in the context of the gauge/gravity duality (also known as AdS/CFT correspondence). The case for astrophysical BHs is nicely summarized in Ref.~\cite{Hughes:2005wj}.

By their very nature, BHs can be seen only indirectly, e.g.~through X-ray spectra of accretion disks surrounding the BH. Therefore, most BHs that have been identified through astrophysical observations are not isolated objects, but have a binary partner --- see for instance the list in Ref.~\cite{Casares:2006vx}, which contains 20 confirmed BHs. At the top of that list is a specific BH, GRS 1915+105 in the constellation Aquila, which has particularly interesting features. 

The mass of GRS 1915+105 is about 18 solar masses, $M=(18.1\pm 0.4)M_\odot$. This experimental result can be established accurately by considering quasi-periodic oscillations (QPOs) \cite{Aschenbach:2004kj}. (Earlier independent measurements led to compatible results, $M=(14\pm 4)M_\odot$, but with higher uncertainty \cite{Greiner:2001,Harlaftis:2003jj}). 
The spin of GRS 1915+105 was measured more recently \cite{McClintock:2006xd,McClintock:2009as}. The extraction of the spin from the observational data is very involved. The method of Ref.~\cite{McClintock:2006xd,McClintock:2009as} employs a spectral analysis of the X-ray continuum, using a suitable accretion disk model to determine the accretion disk parameters and thereby also the BH parameters. The spin, in particular, can be determined very accurately due to its profound impact on the BH properties, such as the innermost stable circular orbit (ISCO). 
The ISCO is the orbit closest to the BH horizon where matter can remain stationary on a circular orbit. The faster the BH spins, the closer the ISCO is to the BH horizon. In the limit of maximal spinning (extremal) BHs the ISCO coincides with the BH horizon. This means that accreting matter in a nearly extremal BH can move much deeper into the gravity well as compared to non-rotating BHs, and this effect leads to a hardening of the X-ray spectrum and a higher efficiency for conversion of accreted rest mass into radiation. Both of these features are observed for GRS 1915+105. It was concluded in Ref.~\cite{McClintock:2006xd} that the dimensionless spin parameter is nearly maximal, $a_\ast>0.98$, close to the Thorne limit $a_\ast<0.998$ \cite{Thorne:1974ve} and to the theoretical upper bound $a_\ast\leq 1$, with $a_\ast=1$ corresponding to an extremal Kerr BH. Therefore, the BH GRS 1915+105 is a nearly extremal Kerr BH.

In view of the importance of accretion disks for the determination of the BH parameters it would be nice to have some exact results available for nearly extremal Kerr BHs. Since back reactions are negligible (see e.g.~\cite{accretion,accretion2}), a suitable accretion disk model involves a fluid on the background of a nearly extremal Kerr geometry. This is still a very hard problem, because viscosity cannot be neglected and electromagnetic fields should be considered as well, cf.~e.g.~\cite{Shakura:1972te,Kluzniak:2000eq,Coppi:2008sa}. A simplification that is useful in other contexts is the Newtonian approximation, which works well if the accretion disk is sufficiently far away from the BH horizon \cite{Shakura:1972te,Kluzniak:2000eq}. However, for nearly extremal BHs accreting matter can come very close to the BH horizon and the Newtonian approximation is not suitable to describe the dynamics close to the ISCO. Thus, if we are interested to describe the dynamics of accreting matter close to the ISCO we have to employ General Relativity. The near-extremality of GRS 1915+105 may allow for a perturbative expansion in the parameter $(1-a_\ast)<0.02$. To leading order in such an expansion the dynamics is described by the extremal limit $a_\ast=1$. We are specifically interested to describe the dynamics close to the ISCO, so we can simultaneously impose a second perturbative expansion, namely a near horizon approximation.

Our starting point is therefore a background geometry that is obtained after a double limit: first, assume that the BH is extremal and second, assume that we are arbitrarily close to the horizon. The geometry obtained in this way is known as Near Horizon Extremal Kerr (NHEK) and was constructed by Bardeen and Horowitz \cite{Bardeen:1999px} (the possibility of viewing the vicinity of the extremal Kerr BH horizon as a spacetime on its own right was already suggested much earlier by the findings of Bardeen and Wagoner \cite{Bardeen:1971}).

In this paper we study exact solutions of perfect and viscous fluids on the NHEK background, and also include a discussion of electromagnetic fields. 

We consider first the standard scenario of a fluid with timelike velocity, $u^a u_a=-1$, which can describe massive as well as massless particles, depending on the equation of state. We construct exact perfect fluid solutions on the NHEK background for a polytropic equation of state. 
However, the Ansatz that $u^a$ be timelike presupposes that there is a local heat bath, which defines a preferred rest frame. Temperature arises here as the zero component of a timelike vector, and transformation to a moving frame introduces a Lorentz factor. If the boost between the heat bath and the reference frame approaches the speed of light the Lorentz factor becomes singular, and one should not consider temperature as the zero component of a timelike vector, but instead consider a lightlike vector. The only natural way to do this is by demanding $u^a u_a=0$.

The dynamics of a fluid in the NHEK geometry is mapped to the dynamics of a fluid at the extremal Kerr BH horizon. Particles moving on stationary orbits at the BH horizon have to move at the speed of light. The same consideration applies to the ``heat bath'', because there is no meaningful way to define a local rest frame at the BH horizon. Everything that remains stationary at the extremal Kerr BH horizon necessarily moves with the speed of light, and the caveat in the previous paragraph applies. Thus, we consider a ``lightlike heat bath'', in the sense that $u^a u_a=0$. We call a fluid with the property $u^a u_a=0$ ``null fluid'' (not to be confused with a timelike fluid, $u^a u_a=-1$, that describes lightlike matter, like a photon fluid). 

We construct a family of exact solutions for a viscous null fluid on the NHEK background in the presence of electromagnetic fields. For a given velocity profile $u^a$ we predict uniquely (up to an overall rescaling) the viscosity function $\eta$. The class of velocity profiles that we encounter has very special properties, and we describe them in detail. 
To address stability issues we discuss linearized perturbations and establish some rigidity results, which point to the linearized stability of our exact solutions.

Besides the potential phenomenological interest related to observations of GRS 1915+100, there are various purely theoretical motivation to consider the NHEK geometry. It is an interesting geometry on its own right and displays several remarkable geometric properties, some of which we review in this work. Uniqueness and stability results were established recently \cite{Amsel:2009et,Amsel:2009ev,Dias:2009ex}. Moreover, the NHEK geometry was exploited in the context of the Kerr/CFT correspondence \cite{Guica:2008mu}, which predicts that any extremal Kerr BH is dual to a certain conformal field theory (CFT). This conjecture has engendered a lot of recent interest (cf.~e.g.~\cite{Lu:2008jk,Azeyanagi:2008kb,Hartman:2008pb,Chow:2008dp,Azeyanagi:2008dk,Peng:2009ty,Ghezelbash:2009gf,Lu:2009gj,Compere:2009dp,Hotta:2009bm,Ghezelbash:2009gy,Krishnan:2009tj,Bredberg:2009pv,Amsel:2009pu,Hartman:2009nz}), and it could be rewarding to apply our results also in this context.

This paper is organized as follows. In section \ref{sec:2} we review salient features of the NHEK geometry and geodesics in its background. In section \ref{sec:3} we construct timelike perfect fluid solutions for arbitrary polytropic equations of state. We mention difficulties with the inclusion of viscosity. In section \ref{sec:4} we construct null perfect fluid solutions. We are able to lift them to viscous null fluid solutions and to include certain electromagnetic fields. In section \ref{sec:5} we address perturbations around our exact solutions. In section \ref{sec:6} we provide a discussion of our results and put them into the perspective of two pertinent branches of literature: BH astrophysics and the Kerr/CFT correspondence.

Before starting we mention some of our conventions. We use signature $-,+,+,+$. We define the Riemann tensor as $R^a{}_{bcd}:=\partial_c\Ga^a{}_{bd}-\partial_d\Ga^a{}_{bc}+\Ga^e{}_{bd}\Ga^a{}_{ce}-\Ga^e{}_{bc}\Ga^a{}_{de}$ and the Ricci tensor as $R_{bd}:=R^a{}_{bad}$, where we employ the usual Einstein summation convention. We choose the sign of the $\epsilon$-tensor so that $\epsilon_{tr\theta\phi}>0$. Symmetrization of indices is defined by $\nabla_{(a}u_{b)}:=\frac12\,(\nabla_a u_b+\nabla_b u_a)$. We employ natural units $c=G=1$.

\section{Near horizon extremal Kerr}
\label{sec:2}

In this section we review the features of Kerr and NHEK most relevant to our work.

\subsection{Kerr geometry and ISCO}

The Kerr metric in Boyer-Lindquist coordinates is
\begin{multline}
ds^2=-\Big(1-\frac{2Mr}{\rho^2}\Big)\,dt^2-\frac{4Mra\sin^2\!\theta}{\rho^2} \,dt\,d\phi\\+\Big(r^2+a^2+\frac{2Mra^2\sin^2\!\theta}{\rho^2}\Big)\sin^2\!\theta\,d\phi^2
+\frac{\rho^2}{\De} \,dr^2+\rho^2\,d\theta^2\,,
\label{eq:kerr}
\end{multline}
with
\eq{
\rho^2 := r^2+a^2\cos^2\!\theta\qquad \De := r^2(1-\frac{2M}{r})+a^2
}{eq:kerrdefs}
It is determined by two physical parameters: the mass $M$ and the angular momentum $J=a M$. Often it is convenient to employ the dimensionless Kerr parameter $a_\ast=a/M$ instead of the angular momentum. With no loss of generality we consider only positive angular momentum, $J,a,a_\ast\geq 0$. The Kerr BH event horizon
\eq{
r_{BH} = M (1 + \sqrt{1-a_\ast^2})
}{eq:NHEK2} 
exists only if the inequality $a_\ast\leq 1$ holds. If it is saturated,
\eq{
a_\ast=1
}{eq:NHEK3}
the Kerr BH becomes extremal. If it is violated, $a_\ast>1$, the Kerr geometry contains a naked singularity. Physical processes like accretion of matter cannot convert a Kerr BH into a naked singularity if the cosmic censorship conjecture is true \cite{Penrose:1999vj}. This suggests that Eq.~\eqref{eq:NHEK3} provides an upper bound for the dimensionless Kerr parameter --- and hence also for the angular momentum in units of the BH mass.\footnote{This point of view was recently challenged by Jacobson and Sotiriou \cite{Jacobson:2009kt} who found that a BH could over-spin by dropping into it a test particle with suitable spin and/or angular momentum, a procedure which requires a careful finetuning. However, as pointed out in their paper, their analysis does neither take into account corrections from gravitational radiation nor from self-force effects, and such effects could spoil the finetuning required for a violation of the cosmic censorship conjecture.}

To determine the ISCO we consider geodesics of timelike test particles in the plane $\theta=\pi/2$, cf.~e.g.~\cite{Chandrasekhar:1985kt}. The radial velocity is determined by 
\eq{
\frac{\dot{r}^2}{2} + V^{\rm eff} = E_N
}{eq:kerr4}
with the effective potential
\eq{
V^{\rm eff}=-\frac{M}{r}+\frac{L^2-2a^2E_N}{2r^2}-\frac{M(L-a\sqrt{2E_N+1})^2}{r^3}
}{eq:kerr5}
Circularity of the orbits requires $V^{\rm eff} = E_N$ and $dV^{\rm eff}/dr=0$. These conditions allow to solve for $E$ and $L$ in terms of $r$. For the ISCO additionally $d^2V^{\rm eff}/dr^2=0$ must hold because this condition separates stable from unstable orbits. This condition allows to solve for the radius of the ISCO $r=r_{\textrm{\tiny ISCO}}$:
\begin{multline}
\frac{r_{\textrm{\tiny ISCO}}^\pm}{M} = 3+\sqrt{x^2+3a^2_\ast} \\
\mp \sqrt{(3+x+2\sqrt{x^2+3a^2_\ast})(3-x)}
\label{eq:kerr8}
\end{multline}
where the upper (lower) sign refers to co-rotation (counter-rotation) and
\eq{
x=1+(1-a^2_\ast)^{1/3}\left[(1+a_\ast)^{1/3}+(1-a_\ast)^{1/3}\right]
}{eq:nolabel}
For $a_\ast\to 0$ we get $x=3$ and the Schwarzschild result $r_{\textrm{\tiny ISCO}}^\pm=6M$ is recovered. In the extremal limit we get $x=1$. The co-rotating ISCO lies at $r_{\textrm{\tiny ISCO}}^+=M$ and thus coincides with the BH horizon $r_{BH}$ in \eqref{eq:NHEK2}:
\eq{
r_{\textrm{\tiny ISCO}}^+ (\textrm{extremal Kerr}) = r_{BH} (\textrm{extremal Kerr}) = M
}{eq:NHEK4}

\subsection{Near horizon extremal Kerr geometry}

If $a_\ast\ll 1$ for many applications it is sufficient to approximate the Kerr solution by the simpler Schwarzschild solution, which is obtained from \eqref{eq:kerr} in the limit $a_\ast\to 0$. Nearly extremal Kerr BHs, $a_\ast\approx 1$, do not allow a very accurate Schwarzschild approximation in the near horizon region close to the ISCO. In that region, however, they allow for a different approximation in terms of the the NHEK geometry introduced by Bardeen and Horowitz \cite{Bardeen:1999px}. The NHEK geometry is obtained from the Kerr geometry \eqref{eq:kerr} as follows. One introduces the dimensionless coordinates
\eq{
\hat t = \frac{\lambda t}{2M}\qquad \hat r = \frac{\lambda M}{r-M}\qquad \hat\phi=\phi-\frac{t}{2M}
}{eq:NHEK5}
and takes the limit $\lambda\to 0$ keeping $\hat t, \hat r, \hat\phi, \theta$ fixed. This yields the NHEK geometry\footnote{In our conventions the relation $J=M^2$ holds.}
\begin{multline}
ds^2=M^2(1+\cos^2\!\theta)\, \Big(\frac{-d\hat t^2+d\hat r^2}{\hat r^2}\\
+\frac{4\sin^2\!\theta}{(1+\cos^2\!\theta)^2}\big(d\hat\phi+\frac{d\hat t}{\hat r}\big)^2+d\theta^2\Big)
\label{eq:NHEK6}
\end{multline}
The spacetime is no longer asymptotically flat --- for instance, at $\theta=0$ the spacetime along the axis is AdS$_2$. Thus, one should not think of the NHEK geometry as an approximation to Kerr in the same way as one considers Schwarzschild as an approximation to Kerr. Rather, the geometry \eqref{eq:NHEK6} describes the (infinite) throat geometry of an extremal Kerr BH, so anything that happens within this geometry is related to physical processes at (or very close to) the event horizon.

The coordinates used in the line element \eqref{eq:NHEK6} do not cover the full NHEK spacetime. Global coordinates, which we shall call again $t, r, \theta, \phi$ (at the minor risk of confusion with the Boyer-Lindquist coordinates), are obtained through the coordinate transformation
\begin{align}
\hat r &= (\sqrt{1+r^2}\cos t+r)^{-1} \\
\hat t &= \hat r \sqrt{1+r^2}\sin t \\
\hat \phi &= \phi + \ln{\Big|\frac{\cos t+r\sin t}{1+\sqrt{1+r^2}\sin t}\Big|}
\end{align}
The NHEK metric in these global coordinates is then given by
\begin{multline}
ds^2=M^2(1+\cos^2\!\theta)\, \Big(-(1+r^2)\,dt^2+\frac{dr^2}{1+r^2}\\
+\frac{4\sin^2\!\theta}{(1+\cos^2\!\theta)^2}\,(d\phi+r\,dt)^2+d\theta^2\Big)
\label{eq:NHEK7}
\end{multline}
The range of the coordinates is $t\in(-\infty,\infty)$, $r\in(-\infty,\infty)$, $\phi\in[0,2\pi)$, $\theta\in[0,\pi]$. Note that $\theta$ maintains its original role as polar angle since it does not appear in any of the coordinate transformations that led from Kerr \eqref{eq:kerr} to NHEK \eqref{eq:NHEK7}. The radial coordinate $r$ is not restricted to be positive because no singularity is encountered at $r=0$. 

\begin{figure}
\epsfig{file=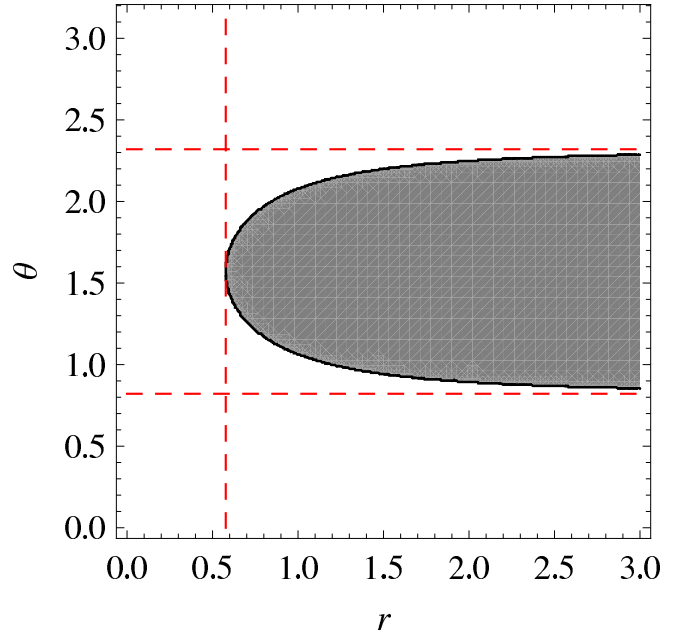, width=\linewidth}
\caption{Norm squared of Killing vector $\partial_t$. The (``ergo''-) region with spacelike $\partial_t$ is plotted in gray, the region with timelike $\partial_t$ is plotted in white. The vertical critical line is at $r=1/\sqrt{3}$. The horizontal critical lines are at $\theta\approx0.82$, $\theta\approx 2.32$.}
\label{fig:1}
\end{figure}

We conclude this brief review with some geometric features of the NHEK spacetime \eqref{eq:NHEK7} and refer to \cite{Bardeen:1999px} for a more complete discussion. The $t=\rm const.$ hypersurfaces are spacelike globally, so there are no closed timelike curves. For sufficiently small radii, $r^2<1/3$ the Killing vector $\partial_\tau$ is timelike. It is also timelike if the inequality $2\sin\theta<1+\cos^2\!\theta$ holds, which is saturated for the critical polar angle $\theta_{\rm crit.}=0.82$ (about $47.1°$). If neither of these inequalities hold $\partial_\tau$ may become spacelike, just like in the ergoregion of the Kerr BH. The plot Fig.~\ref{fig:1} depicts the norm squared of $\partial_\tau$. 
We discuss Killing vectors in more detail in subsection \ref{sec:killing} below. While NHEK does not exhibit any BH horizon, there is still a sense in which one can label a locus in the NHEK geometry with the attribute ``horizon'', namely the horizon of the Poincare-like patch described by the line-element \eqref{eq:NHEK6} at $\hat r\to\infty$.\footnote{The local equivalence between the Poincare-like horizon of NHEK and the event horizon of extremal Kerr has an analog in the simpler case of the extremal Reissner-Nordstr\"om BH and its near horizon limit AdS$_2\times S^2$, see Ref.~\cite{Carroll:2009maa} for a recent discussion. We thank Tom Hartman for discussions on these issues.} If $\lambda$ is finite in \eqref{eq:NHEK5} the limit $\hat r\to\infty$ indeed corresponds to the locus of the horizon. In global coordinates this horizon is mapped to arbitrary finite values of $r$ (if $\hat t$ is infinite) or to $r=\infty$ (if $\hat t$ is finite). Thus, in global coordinates there is not just a single value of the radial coordinate $r$ that would correspond to the original BH horizon, but rather it spreads out through the whole NHEK geometry.

\subsection{Killing vectors, invariants and geodesics}\label{sec:killing}

The Kerr geometry \eqref{eq:kerr} is stationary and axially symmetric. Its Killing vectors
\eq{
k_0 = \partial_t\qquad k_1 = \partial_\phi 
}{eq:NHEK8}
persist in the NHEK geometry \eqref{eq:NHEK7}. The latter has an enhanced isometry group \cite{Bardeen:1999px}, namely $SL(2,\mathbb{R})\times U(1)$, whose algebra is generated by the Killing vectors $k_0, k_1$ and two additional Killing vectors $k_\pm$ defined by
\eq{
k_+ + ik_- = \frac{e^{it}}{\sqrt{1+r^2}} \,\big(r\,\partial_t-i(1+r^2)\partial_r+\partial_\phi\big) 
}{eq:NHEK9}
Generic Killing vectors are neither spacelike nor timelike globally. The only exception is $k_1$, which is spacelike globally.
Scalar fields that are invariant with respect to all four Killing vectors can only depend on the coordinate $\theta$. Vector fields that are invariant with respect to all four Killing vectors must be of the form
\eq{
\big[A_t=r A_\phi(\theta),\;A_r=0,\;A_\theta=A_\theta(\theta),\;A_\phi=A_\phi(\theta)\big]
}{eq:NHEK10}

Since both the Kerr and the NHEK geometry solve the Einstein equations all scalar invariants constructed from the Ricci tensor vanish. There is an interesting scalar invariant that is non-vanishing for the Kerr geometry, namely the Chern-Pontryagin density (also known as instanton density). It is defined by 
\eq{
{\rm CP}:=\frac12\, \epsilon^{cdef}R^a{}_{bef} R^b{}_{acd}
}{eq:NHEK11}
and for the Kerr geometry it is given by
\eq{
{\rm CP} = \frac{96 a M^2 r\cos{\theta}}{(r^2+a^2\cos^2\!\theta)^6} \left(r^2 - 3 a^2\cos^2\!{\theta} \right)\left(3 r^2 - a^2 \cos^2\!{\theta} \right)
}{eq:NHEK12}
The NHEK geometry also has a non-vanishing Chern-Pontryagin density
\eq{
{\rm CP} = \frac{96\cos{\theta}}{M^4(1+\cos^2\!\theta)^6} \left(1 - 3\cos^2\!{\theta} \right)\left(3-\cos^2\!{\theta} \right)
}{eq:NHEK13}
Consistently, the NHEK result \eqref{eq:NHEK13} coincides with the Kerr result \eqref{eq:NHEK12} in the near horizon extremal limit $a=r=M$. In both cases the Chern-Pontryagin density integrates to zero so that the instanton number vanishes. Interestingly, the Chern-Pontryagin density \eqref{eq:NHEK13} of NHEK changes sign at $\theta=\arccos(1/\sqrt{3})\approx 0.955$, an angle we shall encounter again below. As a consistency check we have also compared all other polynomial scalar invariants of NHEK and Kerr in the extremal limit and found that they always coincide.

We review now geodesics of timelike test particles in the plane $\theta=\pi/2$ \cite{Bardeen:1999px}.  The radial velocity is determined by \eqref{eq:kerr4} with the effective potential
\eq{
V^{\rm eff} = \Big(\frac{1}{2M^2}-6L^2\Big)\,r^2 - 8LE\,r
}{eq:NHEK14}
We have normalized the 4-velocity to $u^a u_a=-1$. The constants of motion are given by
\eq{
L = u^\phi + r u^t\qquad E = \frac{1+r^2}{2}\,u^t - 2Lr
}{eq:NHEK15}
The constant $E_N$ on the right hand side of \eqref{eq:kerr4} is determined by $E_N=2E^2-2L^2-1/(2M^2)$. Circularity of the orbits at $r=r_c$ requires that the constants of motion are tuned as follows:
\eq{
r_c=\frac{4EL}{1/(2M^2)-6L^2}\qquad L^2 = \frac{1}{12M^2} 
}{eq:NHEK16}
Therefore, circular orbits are possible only at $|r_c|=\infty$. For $L^2=1/(12M^2)$  the circular orbit is only marginally stable because $d^2V^{\rm eff}/dr^2=0$. An interesting property of timelike geodesics is that they are confined to finite radii if $12L^2<1/M^2$, but can escape to infinity if $12L^2>1/M^2$. The marginal case leads again to the condition $L^2=1/(12M^2)$.

Lightlike geodesics experience a slightly different effective potential. Instead of \eqref{eq:NHEK14} one obtains
\eq{
V^{\rm eff} = - 6L^2\,r^2 - 8LE\,r
}{eq:NHEK17}
The constant $E_N$ on the right hand side of \eqref{eq:kerr4} is now determined by $E_N=2E^2-2L^2$. There are no circular orbits for lightlike test particles because $V^{\rm eff}=E_N$ and $dV^{\rm eff}/dr=0$ cannot hold simultaneously. Lightlike particles moving in the plane $\theta=\pi/2$ generically have a non-vanishing radial component of the velocity.

We generalize now to the case where $\theta$ and $u^\theta$ are arbitrary. The two constants identified in \eqref{eq:NHEK15} take now the form
\begin{align}
& L=\frac{\sin^2\!\theta}{1+\cos^2\!\theta}\, (u^\phi + r u^t) & \label{eq:lgd} \\
& E=\frac{(1+r^2)(1+\cos^2\!\theta)}{2}\,u^t-2 L r &
\end{align} 
Like for the Kerr spacetime \cite{Carter:1968rr} a third constant of motion $\mathcal{D}$ can be identified. For lightlike test particles the corresponding first integral is given by
\eq{
(u^\theta)^2 \,(1+\cos^2\!\theta)^2=\mathcal{D} + 4\cos^2\!\theta \,\Big(L^2-\frac{4 L^2}{\sin^2\!\theta}\Big)
}{eq:geotheta1}
For arbitrary $\theta$ the effective potential (to be inserted into \eqref{eq:kerr4} with $E_N=0$) from \eqref{eq:NHEK17} becomes
\begin{multline}
V^{\rm eff}=2 L^2 (1+r^2)\Big(\frac{1}{(1+\cos^2\!\theta)^2}+\cot^2\!\theta\Big) \\
-\frac{2 (E+2 L r)^2}{(1+\cos^2\!\theta)^2} + \frac{\mathcal{D}(1+r^2)}{2 (1+\cos^2\!\theta)^2}=:V^{\rm eff}_0
\label{eq:geotheta2}
\end{multline}
In the limit $\theta\to\pi/2$, $\mathcal{D}\to 0$ the result \eqref{eq:NHEK17} is recovered, to be inserted into \eqref{eq:kerr4} with $E_N=2E^2-2L^2$.
For timelike test particles the first integral receives one additional contribution as compared to the lightlike case \eqref{eq:geotheta1}:
\eq{
(u^\theta)^2 (1+\cos^2\!\theta)^2=\mathcal{D} + 4\cos^2\!\theta\, \Big(L^2-\frac{1}{4 M^2}-\frac{4 L^2}{\sin^2\!\theta}\Big)
}{eq:geotheta3}
The effective potential for timelike geodesics also contains one additional term as compared to the lightlike case \eqref{eq:geotheta2}:
\begin{equation}
V^{\rm eff}=V^{\rm eff}_0 +\frac{(1+r^2)\sin^2\!\theta}{2 M^2 (1+\cos^2\!\theta)^2}
\label{eq:geotheta4}
\end{equation}
Qualitatively, the potentials \eqref{eq:geotheta2} and \eqref{eq:geotheta4} behave like the simpler potential \eqref{eq:NHEK14}: they contain a term quadratic in $r$ that can change its sign, a term linear in $r$ whose sign is determined by $-LE$ and an $r$-independent term. The discussion about confined geodesics above therefore generalizes straightforwardly to generic geodesics.

We stress that a collection of non-interacting non-rotating ($L=0$) timelike test particles (``dust'') cannot be lifted to a regular perfect fluid solution. This can be seen directly from the effective potential \eqref{eq:NHEK14}, which becomes positive at sufficiently large $r$ if $12L^2<1/M^2$. Consequently, the radial velocity determined from \eqref{eq:kerr4} eventually becomes imaginary. This observation is just a rephrasing of the fact that timelike geodesics with vanishing (or sufficiently small) angular momentum $L$ are confined \cite{Bardeen:1999px}. In the next sections we circumvent this problem by allowing for pressure and find fluid solutions that are well-defined everywhere.

\section{Timelike perfect fluid}
\label{sec:3}

In this section we consider perfect fluid solutions on the NHEK background with a timelike velocity 4-vector $u^a$.

\subsection{Equations of motion}

Without loss of generality we employ the normalization
\eq{
u^au_a = -1
}{eq:NHEK18}
The perfect fluid energy-momentum tensor
\eq{
T^{ab}_{\rm f} = \rho\,u^a u^b + P\,\Delta^{ab}\qquad \Delta^{ab}:=g^{ab}+u^a u^b
}{eq:NHEK19}
must be covariantly conserved
\eq{
\nabla_a T^{ab}_{\rm f} = 0
}{eq:NHEK20}
Here $\rho$ is the density, $P$ is the pressure and both are connected by the polytropic equation of state
\eq{
P=\alpha\, \rho^n
}{eq:NHEK21}
For simplicity we assume for the time being $n=1$.

Contracting the conservation equation \eqref{eq:NHEK20} with $u_b$ leads to the relativistic continuity equation
\eq{
\nabla_a \big(\rho\,u^a\big)+P\,\nabla_au^a=0
}{eq:NHEK22}
Contracting the conservation equation \eqref{eq:NHEK20} with $\Delta^a_b$ leads to the relativistic momentum equations
\eq{
\big(\rho+P\big)\,u^a\nabla_au^b+\Delta^{ab}\,\nabla_a P = 0
}{eq:NHEK23}
These are our equations of motion. The simple structure of the NHEK background \eqref{eq:NHEK7} considerably reduces the amount of terms contributing to the equations of motion. Moreover, by analogy to the Kerr case \cite{Novikov:1973,Riffert:1995,Riffert:2000,Abramowicz:1996ap} we shall exclusively consider velocity profiles that are stationary and axisymmetric.
\eq{
u^a=u^a(r,\theta)
}{eq:NHEK56}
In appendix \ref{app:A} the equations of motion are displayed explicitly for arbitrary velocity profiles of the form above \eqref{eq:NHEK56}. For vanishing polar velocity $u^\theta=0$ the continuity equation \eqref{eq:NHEK22} evaluated on the NHEK background \eqref{eq:NHEK7} reduces to
\eq{
u^r\partial_r\rho+(\rho+P)\,\partial_ru^r=0\qquad[u^\theta=0]
}{eq:NHEK24}

\subsection{Exact solutions}

We consider now solutions to the equations of motion with vanishing polar velocity $u^\theta=0$.
A simple class of such solutions can be obtained by assuming $u^r=0$, which obviously solves the continuity equation \eqref{eq:NHEK24}. 
Since the equations of motion are still somewhat lengthy we restrict our attention to the plane $\theta=\pi/2$ for the time being. In additions to the equations of motion the normalization condition \eqref{eq:NHEK18} must be fulfilled. The simplest way to achieve the latter is by demanding $u_\phi=0$
and therefore
\eq{
u^t = \frac{1}{M\sqrt{(1+r^2)}}\qquad u^\phi=-r\,u^t
}{eq:NHEK25}
Physically, this velocity profile corresponds to choosing vanishing angular momentum of the fluid particles, see the left Eq.~\eqref{eq:NHEK15}. The continuity equation holds per Ansatz, and the momentum equations evaluated at $\theta=\pi/2$ lead to only two conditions
\begin{align}
& \partial_\theta \rho =0 & [\theta=\pi/2] & \\
& \partial_r \rho +\frac{(1+\al)\,r}{\al\,(1+r^2)}\,\rho=0 & [\theta=\pi/2] &
\end{align}
The momentum equations are solved by the following density and momentum profiles at $\theta=\pi/2$:
\eq{
\rho(r,\theta=\pi/2) = \rho_0 \,(1+r^2)^{-\frac{1+\alpha}{2\alpha}}\qquad P=\alpha\,\rho
}{eq:NHEK26}

Our next goal is to lift the solution in the plane $\theta=\pi/2$ given by \eqref{eq:NHEK25}, \eqref{eq:NHEK26} to a solution valid in the whole NHEK spacetime. This is indeed possible and our final result is
\begin{subequations}
\label{eq:pfsol}
\begin{align}
u^t &= \frac{1}{M\sqrt{(1+r^2)(1+\cos^2\!\theta)}}\\
u^r &= 0 \\
u^\theta &= 0\\
u^\phi &= -\frac{r}{M\sqrt{(1+r^2)(1+\cos^2\!\theta)}}\\
\rho &= \rho_0 \,\big((1+r^2)(1+\cos^2\!\theta)\big)^{-\frac{1+\alpha}{2\alpha}} \\
P &= \alpha \,\rho
\end{align}
\end{subequations}
The configuration \eqref{eq:pfsol} solves the equations of motion \eqref{eq:NHEK22}, \eqref{eq:NHEK23}.
It is gratifying that reasonable values of the equation of state parameter, $\alpha\in[-1,1]$, always lead to a density profile that is finite in the whole spacetime. Note also that there are no divergences induced by the dependence on the polar angle. To the best of our knowledge \eqref{eq:pfsol} is the first exact perfect fluid solution constructed on the NHEK background \eqref{eq:NHEK7}.

We generalize now the solution \eqref{eq:NHEK26} to the polytropic equation of state \eqref{eq:NHEK21} with $n\neq 1$ (and $n>0$). By solving the momentum equations in the plane $\theta=\pi/2$ we obtain a density profile of the form
\eq{
\rho(r,\theta=\pi/2)= \Big(-\frac{1}{\al}+\rho_0\,(1+r^2)^{\frac{1/n-1}{2}}\Big)^{\frac{1}{n-1}} 
}{eq:NHEK42}
and $P=\al\,\rho^n$. For negative $\alpha$ the solution is always globally well-defined. For positive $\alpha$ and $n>1$ the density inevitably becomes negative in some region of spacetime and such solutions must be discarded. For positive $\alpha$ and $0<n<1$ the solution is globally well-defined provided that the inequality $\alpha \rho_0> 1$ holds. 

The thickness of an accretion disk is usually derived from a balance between geodesic deviation (gravitational forces) and other physical forces (see for instance the appendix of \cite{McClintock:2006xd}). We calculate now the geodesic deviation  between two nearby test particles. Given a distance vector $x^a$ the acceleration $g^a$ between these particles can be calculated from the geodesic deviation equation
\eq{
g^a = R^a{}_{bcd} u^b u^c x^d
}{eq:geodev1}
where $u^a$ is the 4-velocity. For the solution \eqref{eq:pfsol} and for a vector $x^\theta\neq 0$, $x^a=0$ otherwise, we obtain
\eq{
g^\theta = x^\theta \frac{1-3\cos^2\!\theta}{M^2\,(1+\cos^2\!\theta)^3}
}{eq:geodev2}
Interestingly, there is a zero in the acceleration for a certain polar angle, even without the presence of additional forces. The critical value is given by $\theta\approx 0.955$ (about 54.74 degrees). Curiously, this value of $\theta$ also leads to a vanishing Chern-Pontryagin density \eqref{eq:NHEK13}.

We do not further dwell on the perfect fluid case since our main interest are viscous fluids.

\subsection{Shear viscosity}

In the presence of shear viscosity the energy momentum tensor for a viscous relativistic fluid receives an additive contribution
\eq{
T^{ab}_{\rm vf} = T^{ab}_{\rm f} +\Pi^{ab}
}{eq:NHEK27}
The viscosity tensor $\Pi^{ab}$ is given by
\eq{
\Pi^{ab} = -2\eta\, \big(\De^{ac}\De^{bd}\,\nabla_{(c}u_{d)}-\frac13\,\De^{ab}\,\nabla_cu^c\big)
}{eq:NHEK28}
The function $\eta$ parameterizes the local strength of viscosity. Three important properties of the viscosity tensor are symmetry, tracelessness and transversality:
\eq{
\Pi^{ab}=\Pi^{ba}\qquad \Pi^a{}_a=0\qquad \Pi^a{}_b u^b = 0
}{eq:NHEK29}
The tracelessness is a consequence of neglecting bulk viscosity. The continuity equation generalizing \eqref{eq:NHEK22} is now given by
\eq{
\nabla_a \big(\rho\,u^a\big)+P\,\nabla_au^a+\Pi^{ab}\,\nabla_a u_b = 0
}{eq:NHEK30}
The momentum equations generalizing \eqref{eq:NHEK23} are now given by
\eq{
\big(\rho+P\big)\,u^a\nabla_au^b+\Delta^{ab}\,\big(\nabla_a P+\nabla_c\Pi^c{}_a\big) = 0
}{eq:NHEK31}
In the Newtonian limit only the momentum equations receive corrections from viscosity. It may be checked easily that the momentum equations \eqref{eq:NHEK31} are fulfilled for our solution \eqref{eq:pfsol}, provided the viscosity function is chosen as
\eq{
\eta = \frac{\eta_0(\theta)}{\sqrt{1+r^2}}\qquad[\textrm{solves momentum equations}]
}{eq:NHEK32}
Here $\eta_0$ is an arbitrary function of the polar angle $\theta$. Technically, the reason for the persistence of the solution \eqref{eq:pfsol} with \eqref{eq:NHEK32} at the level of momentum equations \eqref{eq:NHEK31} is that for $u^r=u^\theta=0$ the only non-vanishing components of the viscosity tensor are $\Pi^{r\phi}$ and $\Pi^{tr}$ (actually, the latter component also vanishes in our case). However, the continuity equation \eqref{eq:NHEK30} is violated unless $\eta_0$ vanishes. Thus there is no simple way to lift our perfect fluid solution \eqref{eq:pfsol} to a viscous fluid solution.

It is important to understand why this happens, so that a more refined analysis can resolve this problem. As we have pointed out, the perfect fluid continuity equation \eqref{eq:NHEK22} holds identically for $u^r=u^\theta=0$. Therefore, the contribution from the viscosity tensor to the continuity equation would have to vanish by itself. This is not the case for the configuration \eqref{eq:pfsol} with \eqref{eq:NHEK32} unless $\eta_0=0$. We conclude that a configuration for a viscous fluid should either contain a radial velocity component, $u^r\neq 0$, or a polar velocity component, $u^\theta\neq 0$, or both, unless the identity $(\nabla_{(a} u_{b)})(\nabla^au^b)=0$ holds. We have not succeeded in finding exact solutions with viscosity for the timelike case. However, we are able to construct solutions for the lightlike case discussed below.

\section{Viscous null fluid}
\label{sec:4}

In this section we consider lightlike viscous fluids 
\eq{
u^a u_a=0
}{eq:NHEK57}
and construct exact solutions. We also add an electromagnetic field.

\subsection{Physical preliminaries}\label{sec:prel}

Before writing down any equations we want to include some input from physics, because this will guide us to finding appropriate solutions of the equations of motion. First of all, as explained in the introduction, the lightlike case is relevant for physics at or near the ISCO. Second, it is a reasonable approximation to consider vanishing angular momentum, because frame dragging effects acting on particles near the ISCO dominate over the relative movement on top of the frame dragging effects. From the left Eq.~\eqref{eq:NHEK15} we may therefore assume that $u^\phi = -r\,u^t$. Third, we want to include viscosity and have experienced difficulties with such an inclusion if $u^r=u^\theta=0$ (see previous section). Therefore, we should refrain from setting both of these velocity components to zero. In addition to these physically motivated conditions we assume for the time being the polar velocity to be negligible as compared to the other velocity components. Consequently, we employ again the simplification $u^\theta=0$. The only unknown functions in the 4-velocity are therefore $u^t$ and $u^r$, and they are related by the null condition $u^a u_a=0$. Putting all these ingredients together leads to the following Ansatz for the 4-velocity:
\eq{
u^t=u^t(r,\theta)\quad u^r = (1+r^2)\,u^t\quad u^\theta=0\quad u^\phi=-r\,u^t
}{eq:NHEK33}
Additional constraints come from our desire to eventually match scalar quantities in the NHEK geometry with corresponding quantities in the Kerr geometry evaluated at/near the ISCO. However, we have seen in section \ref{sec:killing} that the ISCO at $r=M$ in the Kerr geometry is mapped to arbitrary values of the radial coordinate $r$ in the NHEK geometry. Thus, a sensible matching is possible only if the corresponding scalar quantities are independent from $r$. Since we also require $\phi$ and $t$ independence for our background solution, all scalars in the NHEK geometry that allow for a sensible matching must be invariant with respect to all four Killing vectors \eqref{eq:NHEK8}, \eqref{eq:NHEK9} --- they can only depend on the polar angle $\theta$. Thus, we assume
\eq{
\rho=\rho(\theta)\qquad P=\frac13\,\rho\qquad \eta=\eta(\theta)
}{eq:NHEK34}
where for consistency we have imposed the relativistic equation of state for a lightlike fluid.

The energy momentum tensor of a viscous null fluid is given by
\eq{
T^{ab}_{\rm vnf} = P \big(4u^a u^b + g^{ab}\big) + \Pi^{ab}
}{eq:NHEK47}
The shear viscosity tensor is slightly different as compared to the timelike case \eqref{eq:NHEK28}:
\eq{
\Pi^{ab} = -2\eta\, \big(\De^{ac}\De^{bd}\,\nabla_{(c}u_{d)}-\frac14\,\De^{ab}\,\nabla_cu^c\big)
}{eq:NHEK44}
Instead of the properties \eqref{eq:NHEK29} we obtain
\eq{
\Pi^{ab}=\Pi^{ba}\qquad \Pi^a{}_a=0\qquad \Pi_{ab}u^a u^b=0
}{eq:NHEK45}
but {\em not} necessarily $\Pi^a{}_b u^b=0$. The continuity equation in general simplifies to
\eq{
u^a\partial_a P + u_a \nabla_b \Pi^{ab} = 0
}{eq:NHEK46}
Together with the Ansatz $u^\theta=0$, $P=P(\theta)$, the continuity equation \eqref{eq:NHEK46} turns into a constraint on the viscosity tensor.
\eq{
u_a \nabla_b \Pi^{ab} = 0
}{eq:NHEK49}
The momentum equations can be presented as
\eq{
\big(4u^a u^b + g^{ab}\big)\,\partial_a P + 4P\,\nabla_a(u^a u^b)  + \nabla_a \Pi^{ab} = 0
}{eq:NHEK48}
Of course, only three of them are independent from the continuity equation \eqref{eq:NHEK46}.

\subsection{Exact solutions}\label{sec:exact}

Our task is now to find an appropriate velocity profile $u^t(r,\theta)$ and expressions for $\rho(\theta)$ and $\eta(\theta)$. The derivations are presented in appendix B. 
Thus, given the Ansatz \eqref{eq:NHEK33}, \eqref{eq:NHEK34} we can present already the most general exact relativistic viscous fluid solution in the NHEK background:
\begin{subequations}
\label{eq:vfsol}
\begin{align}
u^t &= \frac{u_0(\theta)}{(1+r^2)(1+\cos^2\!\theta)} \label{eq:utsol} \\
u^r &= \frac{u_0(\theta)}{1+\cos^2\!\theta} \\
u^\theta &= 0\\
u^\phi &= -\frac{r\,u_0(\theta)}{(1+r^2)(1+\cos^2\!\theta)}\\
\rho &= \rho_0 \\
P &= \frac13 \,\rho_0\\
\eta &= \eta_0(\theta)
\end{align} 
The functions $\eta_0$ and $u_0$ are not independent from each other but related by the differential equation
\begin{multline}
 \eta'_0\,\Big( u_0' +\frac{2\sin\theta\cos\theta}{1+\cos^2\!\theta}\, u_0\Big) \\
+ \eta_0\,\Big( u_0'' + \frac{2 \cot\theta + \sin\theta\cos\theta}{1+\cos^2\!\theta}\, u_0'\Big)=0
\label{eq:NHEK43}
\end{multline}
\end{subequations}
An interesting property of our solution is transversality with respect to velocity
\eq{
\Pi^{ab}u_b=0
}{eq:NHEK58}
Thus, even though the viscosity tensor \eqref{eq:NHEK44} in general does not exhibit the transversality property \eqref{eq:NHEK58}, it has this property for the most general solution \eqref{eq:vfsol} compatible with the Ansatz \eqref{eq:NHEK33}, \eqref{eq:NHEK34}. Therefore, our viscosity tensor has all the standard properties \eqref{eq:NHEK29} like for an ordinary fluid in the Landau-Lifshitz frame \cite{lali:fluid}. Other useful properties of the viscosity tensor and the velocity profile \eqref{eq:vfsol} are the following ones:
\begin{subequations}
\label{eq:ident}
\begin {align}
&        \nabla_a \Pi^{ab}=0 \label{eq:id1} \\ 
&	\nabla_a u^a=0 \label{eq:id2} \\
&	u^a \nabla_a u^b=0 \label{eq:id3} \\
&        \nabla_a u_b - \nabla_b u_a = 0 \qquad \Leftrightarrow \qquad u_0=\rm const. \label{eq:id5}
\end{align}
\end{subequations}
Thus, the viscosity tensor and the velocity are divergence-free \eqref{eq:id1}, \eqref{eq:id2}. The velocity profile is geodetic \eqref{eq:id3}. It has vanishing twist (vorticity) if and only if $u_0$ is constant \eqref{eq:id5}. In that case the velocity profile coincides with the velocity profile of lightlike geodesics with $L=0$ in \eqref{eq:lgd}-\eqref{eq:geotheta2}. The shear of the velocity profile is non-vanishing in general, $\nabla_{(a}u_{b)}\neq 0$, but its norm vanishes, $\nabla_{(a}u_{b)} \nabla^{(a}u^{b)}=0$, as a direct consequence of the properties \eqref{eq:id2}, \eqref{eq:id3}.

If the velocity profile is known (or can be guessed on physical grounds) the differential equation \eqref{eq:NHEK43} leads to a prediction for the viscosity profile. Of course, not all solutions that are allowed mathematically make sense physically --- the velocity and/or viscosity profiles may have singularities at certain angles, or viscosity may fail to be non-negative in the whole spacetime. The simplest solution \eqref{eq:vfsol} that is physically meaningful in the whole NHEK spacetime has constant $u_0$ and constant viscosity $\eta_0>0$. According to \eqref{eq:id5} this is the only solution with vanishing vorticity. A solution with vorticity is given by $u_0\propto \cos\theta+\ln{\tan^2\frac{\theta}{2}}$ and constant viscosity $\eta_0>0$. This solution has a velocity profile with a logarithmic singularity at the poles $\theta=0,\pi$, but is regular and physically acceptable otherwise.

\subsection{Conserved currents}\label{sec:cons}

We construct now conserved currents by following a standard procedure. We start with the equations of motion \eqref{eq:NHEK48} and contract them with the Killing vectors \eqref{eq:NHEK8}, \eqref{eq:NHEK9}. By virtue of the Killing equation $\nabla_{(a}k_{b)\,0,1,\pm}=0$ we obtain currents
\eq{
J^a_{0,1,\pm} = P\,k^a_{0,1,\pm} +4 u^a u_bk^b_{0,1,\pm} +  \Pi^a_b\,k^b_{0,1,\pm} 
}{eq:NHEK400}
that are conserved
\eq{
\nabla_a J^a_{0,1,\pm} = 0
}{eq:NHEK401}
We discuss now briefly some properties of these currents evaluated on the solutions \eqref{eq:vfsol}.
The current related to energy flux is given by
\eq{
J^a_0 =  P\de^a_t + 4 u^a u_t + \Pi^a_t 
}{eq:NHEK402}
Depending on the velocity profile function $u_0(\theta)$ this current may be spacelike or timelike in certain regions of spacetime. For large values of $|r|$ the current \eqref{eq:NHEK402} is spacelike between the two horizontal asymptotes depicted in Fig.~\ref{fig:1}, essentially the ``ergoregion'', and timelike elsewhere, regardless of the velocity profile. The current related to angular momentum flux is given by
\eq{
J^a_1 = P\de^a_\phi + \Pi^a_\phi 
}{eq:NHEK403}
For any velocity profile function $u_0(\theta)$ this current is spacelike everywhere, except at the poles $\theta=0,\pi$ where it becomes lightlike. This result can be seen easily by exploiting the property
\eq{
\Pi_{ab}\,\Pi^b{}_c = 0 \qquad \textrm{if\;\,} a=\theta,\phi \textrm{\quad or\;\,} c=\theta,\phi
}{eq:NHEK404}
which implies $J^a_1 J_{a\,1}\propto k^a_1 k_{a\,1}\propto g_{\phi\phi}$. The remaining two currents have no analogue in the Kerr spacetime, and we have not found a simple physical interpretation for them.

Integrating the current conservation equation \eqref{eq:NHEK401} over some spacetime volume $V$ and using Gauss' law leads to integral identities
\eq{
\int_{\partial V}\!\!\! d^3x\,\sqrt{|\ga|}\, J^a_{0,1,\pm} n_a = 0
}{eq:NHEK406}
Here $\partial V$ is the boundary of the spacetime volume, $n^a$ the outward pointing unit normal and $\ga$ the determinant of the induced metric at the boundary. Of particular interest are boundaries at constant radius, $r=r_0$, so that $n^a = \sqrt{g^{rr}}\,(\partial_r)^a$ and $\sqrt{|\ga|}=\sqrt{-g}\,\sqrt{g^{rr}}$. If we consider as volume $V$ an interval $r\in[r_0,r_0+\eps]$, with the other coordinates arbitrary, then the integral identity \eqref{eq:NHEK406} in the limit of $\eps\to 0$ simplifies to\footnote{Note that the determinant of the metric is independent from $r$.}
\eq{
\int_{r=r_0}\!\!\!\!\!\! d^3x\, \sqrt{-g}\,\partial_r J^r_{0,1,\pm} = 0
}{eq:NHEK407a}
Thus, if the local identity
\eq{
\partial_r \,J^r_{0,1,\pm} = 0
}{eq:NHEK411}
holds the integral identity \eqref{eq:NHEK407a} is implied automatically. 

A particularly simple case is the integral identity \eqref{eq:NHEK407a} for the angular momentum flux $J_1$. Dropping the trivial $t$ and $\phi$ integrations we obtain
\eq{
\int\limits_0^\pi d\theta\,\sin\theta\,(1+\cos^2\!\theta)\,\partial_r \Pi^r_\phi
=  0
}{eq:NHEK408}
With the formula for $\Pi_{r\phi}$ from the appendix \eqref{eq:appeq} and the solution for $u^t$ \eqref{eq:utsol} we obtain
\eq{
\Pi^r_\phi =  \frac{4\eta_0 u_0\,\sin^2\!\theta}{(1+\cos^2\!\theta)^3} \geq 0
}{eq:NHEK409}
Therefore, the stronger condition \eqref{eq:NHEK411} is fulfilled and the integral identity \eqref{eq:NHEK408} holds trivially. The inequality in Eq.~\eqref{eq:NHEK409} is true for physical reasons:  both the viscosity function $\eta_0$ and the time component of the velocity (and hence $u_0$) must be non-negative. 

We postpone applications of the integral identities \eqref{eq:NHEK407a} and a comparison to analog identities in the Kerr case to the conclusions.

\subsection{Electromagnetic field}

To describe a relativistic viscous plasma we introduce the abelian field strength
\eq{
F_{ab} = \nabla_a A_b - \nabla_b A_a = \partial_a A_b - \partial_b A_a
}{eq:NHEK36}
in terms of the 4-vector potential $A_a$. Then the homogeneous Maxwell equations
\eq{
\epsilon^{abcd}\,\nabla_b F_{cd} = 0
}{eq:NHEK37}
hold automatically. The inhomogeneous Maxwell equations
\eq{
\nabla_a F^{ab} = 4\pi\,j^b
}{eq:NHEK38}
are sourced by the 4-current $j^a$ (with some convenient normalization). The full energy momentum tensor
\eq{
T^{ab}_{\rm vnp} = T^{ab}_{\rm vnf} + T^{ab}_{\rm M}
}{eq:NHEK39}
receives a contribution from the Maxwell field:
\eq{
T^{ab}_{\rm M} = \frac{1}{4\pi}\,\Big(F^{ac}F^b{}_c-\frac14\,g^{ab}\,F^{cd}F_{cd}\Big)
}{eq:NHEK40}
We consider here exclusively solutions where the Maxwell energy momentum tensor is conserved by itself.
\eq{
\nabla_a T^{ab}_{\rm M} = 0
}{eq:NHEK41}

We have found several solutions with vanishing or non-vanishing current. The latter all turned out to be problematic in the following sense: there is always a part of the spacetime where the current becomes spacelike, $j^a j_a > 0$. This need not be a generic feature, but it is a feature present in all the solutions we have found. Therefore, we consider only solutions with $j^a=0$. 

A particular solution of this type is given by the gauge field
\begin{subequations}
\label{eq:NHEK55}
\begin{align}
A_t &= B\,\ln{(1+r^2)} - E\,\big(\cos\theta+\ln\tan^2\!\frac{\theta}{2}\big)\\
A_\phi &= 2B\,\arctan{r}\\
A_r &= 0 \\
A_\theta &= 0
\end{align}
\end{subequations} 
The field strength can be decomposed into electric and magnetic parts with respect to a 4-vector $n^a$, which is usually assumed to be timelike and normalized. Since we do not have some preferred timelike vector field available, we choose instead $n^a=u^a$. The associated magnetic field is given by $B_a = \frac12\,n^b \epsilon_{bacd}\,F^{cd}$ with the non-vanishing components
\eq{
B^\theta = -\frac{B\,u^t}{M^2\,\sin\theta}\qquad B^\phi= \frac{E\,u^t (1+\cos^2\!\theta)}{2M^2\,\sin^2\!\theta}
}{eq:NHEK50}
The associated electric field is given by $E_a = n^b F_{ba}$, with the non-vanishing components
\eq{
E^\theta=\frac{E\,u^t}{M^2\,\sin\theta}\qquad E^\phi = \frac{B\,u^t (1+\cos^2\!\theta)}{2M^2\,\sin^2\!\theta} 
}{eq:NHEK51}
The action invariant $-\frac14\,F_{ab}F^{ab}$ for the solution \eqref{eq:NHEK55} is given by
\eq{
-\frac14\,F_{ab}F^{ab} = \frac{E^2-B^2}{2M^4\,(1+r^2)\,\sin^2\!\theta}
}{eq:NHEK52}
The instanton invariant $\epsilon_{abcd}\,F^{ab}F^{cd}$ for the solution \eqref{eq:NHEK55} is given by 
\eq{
\epsilon_{abcd}\,F^{ab}F^{cd} = \frac{8EB}{M^2\,(1+r^2)\,\sin^2\!\theta}
}{eq:NHEK53}
Both invariants exhibit second order poles at the two poles $\theta=0,\pi$. The action invariant is zero if $E=\pm B$, the instanton invariant is zero if either $E$ or $B$ vanishes. 

Our discussion of electromagnetic fields is far from being exhaustive, but it clearly demonstrates that non-vanishing electromagnetic fields can be included straightforwardly.

\section{Perturbations}
\label{sec:5}

In this section we consider small deviations from stationarity and axisymmetry. We decompose all fields into a background contribution and a fluctuation. Our main assumption is that the fluctuations in the viscosity function $\eta$ are negligible as compared to fluctuations in density or fluctuations in velocity, concurrent with previous approaches (cf.~e.g.~\cite{Rebusco:2009ui}). We also assume that fluctuations do not change the equation of state, $P=\rho/3$. Thus, we make the Ansatz
\begin{subequations}
\label{eq:pert1}
\begin{align}
u^a_f &= u^a + \epsilon \,\delta u^a\\
\rho_f &= \rho + \epsilon\, \delta \rho \\
P_f &= P + \epsilon\, \delta P = \frac13\,\rho_f = \frac13\,\rho + \epsilon\, \frac13\,\delta \rho\\
\eta_f &= \eta 
\end{align}
\end{subequations}
Quantities with an index $f$ contain both the background solution (denoted without index) and the fluctuation (denoted by $\delta$ and scaled by the small parameter $\epsilon$). The fluctuation functions $\delta u^a$ and $\delta\rho$ depend on all coordinates $t$, $r$, $\theta$, $\phi$. We demand that the total velocity be lightlike.
\eq{
u^a_f u_a^f =  2 \epsilon \,u_a \delta u^a + O(\epsilon^2) = 0
}{eq:NHEK85}
The inner product between the background velocity $u^a$ and the velocity fluctuation vector $\delta u^a$ must therefore vanish to leading order in $\epsilon$. In this work we shall exclusively consider first order expressions in $\epsilon$, and many of the equalities below are valid not exactly, but only up to higher order terms. Another assumption concerns the transversality property \eqref{eq:NHEK58}: we demand that this property is maintained to leading order in the fluctuations,
\eq{
u^b\,\de\Pi_b^a+\Pi_b^a\, \de u^b = 0 
}{eq:NHEK111}
The property \eqref{eq:NHEK111} guarantees that also at linearized order in the fluctuations our viscosity tensor has all the standard properties \eqref{eq:NHEK29}. For similar reasons as in section \ref{sec:prel} we require that scalar quantities like pressure exhibit no radial dependence,
\eq{
\partial_r \de P=0
}{eq:NHEK112}

We assume further that the fluctuations do not introduce angular momentum,
\eq{
\de u_\phi = 0
}{eq:NHEK83}
The conditions \eqref{eq:NHEK85} and \eqref{eq:NHEK83} imply that the velocity fluctuation vector has only two independent components and conveniently can be decomposed as
\eq{
\de u^a = u^a \de u + \de^a_\theta \de u^\theta
}{eq:NHEK84}
The first contribution to the velocity fluctuation vector \eqref{eq:NHEK84} leads to a rescaling of the velocity. We call these fluctuations ``scaling fluctuations''. The second contribution gives the velocity profile a $\theta$-component and thus provides a vertical component to the velocity vector in cylindrical coordinates. We call these fluctuations ``vertical fluctuations''. We shall discuss both contributions separately.

Before such a discussion we address an important technical issue. Below we shall encounter various partial differential equations of the form
\eq{
u^a\partial_a f = 0 \quad \Leftrightarrow\quad (\partial_t+(1+r^2)\partial_r-r\partial_\phi) f = 0
}{eq:NHEK87}
where $f$ is some physical quantity and $u^a$ is the background velocity profile given by \eqref{eq:vfsol}. The general solution of the differential equation \eqref{eq:NHEK87} consistent with periodicity in the azimuthal angle $\phi$ is given by
\begin{multline}
f = \sum_{n=0}^\infty f^c_n(t-\arctan{r},\theta)\,\cos{\big(n(\phi+\frac12\,\ln{(1+r^2)})\big)} \\
+ \sum_{n=1}^\infty f^s_n(t-\arctan{r},\theta)\,\sin{\big(n(\phi+\frac12\,\ln{(1+r^2)})\big)}
\label{eq:NHEK88}
\end{multline}
The free functions $f^c_n$, $f^s_n$ depend on two arguments: the first one is the combination $(t-\arctan{r})$ and the second the polar angle $\theta$.

\subsection{Vertical fluctuations}

We set $\de u=0$ and keep only $\de u^\theta$ in the velocity fluctuation vector \eqref{eq:NHEK84}. 

The $\theta$-component of the transversality condition \eqref{eq:NHEK111} implies
\eq{
\eta\, u^c \partial_c \de u^\theta = 0
}{eq:NHEK300}
Therefore, $\de u^\theta$ must have the same Fourier decomposition as the function $f$ in \eqref{eq:NHEK88}. The $\phi$-component of the transversality condition \eqref{eq:NHEK111} holds identically. The $t$- and $r$-components of the transversality condition \eqref{eq:NHEK111} are redundant with each other and yield another condition.
\eq{
u^a\,\de\Pi^r_a + \Pi^r_\theta \,\de u^\theta = 0
}{eq:NHEK301}
With the definition \eqref{eq:NHEK44} for the viscosity tensor the condition \eqref{eq:NHEK301} establishes a first order differential equation for the fluctuation $\de u^\theta$. 
\eq{
\partial_\theta \,\de u^\theta = \Big(\frac{2u_0^\prime}{u_0}+\frac{(1-3\cos^2\!\theta)\cot\theta}{1+\cos^2\!\theta}\Big)\,\de u^\theta
}{eq:NHEK303}
Its general solution is given by
\eq{
\de u^\theta = \frac{u_0^2(\theta)}{\sin\theta\,(1+\cos^2\!\theta)}\,f(t,r,\phi)
}{eq:NHEK302}
Of course, the integration function $f(t,r,\phi)$ is not arbitrary, but must have the Fourier decomposition \eqref{eq:NHEK88}.
For any background profile that has non-vanishing $u_0$ at the poles $\theta=0,\pi$, the solution \eqref{eq:NHEK302} necessarily diverges there. If $\de u^\theta$ diverges at some points it should not be regarded as a small (first order) fluctuation. Therefore, without having to solve any of the equations of motion we are led to a rigidity result: there are no vertical fluctuations as long as the background velocity does not vanish at the poles $\theta=0,\pi$. This applies to the cases $u_0=\rm const.$ and $u_0\propto\cos\theta+\ln{\tan^2\frac{\theta}{2}}$ discussed in section \ref{sec:exact}. Vertical fluctuations with a regular profile for $\de u^\theta$ can exist only if $u_0(\theta)$ vanishes at the poles $\theta=0,\pi$.

\subsection{Scaling fluctuations}

We set $\de u^\theta=0$ and keep only $\de u$ in the velocity fluctuation vector \eqref{eq:NHEK84}. 

The continuity equation \eqref{eq:NHEK46} simplifies to 
\eq{
u^a\partial_a \de P + u_a \nabla_b\de\Pi^{ab} = 0
}{eq:NHEK200}
Transversality \eqref{eq:NHEK111} implies 
\eq{
u_a \de\Pi^{ab} = -\frac{\eta}{2}\,u^b u^c\partial_c \de u = 0
}{eq:NHEK203}
It can be shown that transversality also implies $u_a \nabla_b\de\Pi^{ab}=0$. Therefore, the continuity equation further reduces to
\eq{
u^a\partial_a \de P = 0
}{eq:NHEK201}
Together with the requirement \eqref{eq:NHEK112} we obtain the following result from the general solution \eqref{eq:NHEK88}:
\eq{
\de P = \de P (\theta)
}{eq:NHEK202}
Similarly, from the condition \eqref{eq:NHEK203} we obtain
\begin{multline}
\de u = \sum_{n=0}^\infty \de u^c_n(t-\arctan{r},\theta)\,\cos{\big(n(\phi+\frac12\,\ln{(1+r^2)})\big)} \\
+ \sum_{n=1}^\infty \de u^s_n(t-\arctan{r},\theta)\,\sin{\big(n(\phi+\frac12\,\ln{(1+r^2)})\big)}
\label{eq:NHEK204}
\end{multline}

The momentum equation \eqref{eq:NHEK48} with free index $\phi$ simplifies to
\eq{
\nabla_a\de \Pi^a_\phi = 0
}{eq:NHEK100}
and is fulfilled identically. The momentum equation \eqref{eq:NHEK48} with free index $\theta$ then simplifies to
\eq{
\partial_\theta \de P = 0
}{eq:NHEK104}
which implies that $\de P=\rm const.$ A constant shift of the constant background pressure can be absorbed by a redefinition of units and therefore no physical pressure fluctuations are encountered in scaling fluctuations. The velocity fluctuations, however, can be non-trivial.

The remaining two momentum equations with free indices $t,r$ are redundant with each other. We consider the $r$ equation.
\eq{
\nabla_a \de \Pi^a_r = 0
}{eq:NHEK105}
It leads to a second order differential equation in the coordinate $\theta$. Since we can linearly superpose fluctuations it is sufficient to consider one of the Fourier modes. We take
\eq{
\de u = f_n^c(t-\arctan{r},\theta)\,\cos{\big(n(\phi+\frac12\,\ln{(1+r^2)})\big)}
}{eq:NHEK206}
for some fixed integer $n$. The second order differential equation descending from the $r$-momentum equation \eqref{eq:NHEK105} reads explicitly
\begin{multline}
\Big(\partial_\theta^2+\big(\frac{\eta_0^\prime}{\eta_0}+\frac{2u_0^\prime}{u_0}+\frac{(3-\cos^2\!\theta)\cot\theta}{1+\cos^2\!\theta}\big)\,\partial_\theta\\
-n^2\,\frac{(1+\cos^2\!\theta)^2}{4\sin^2\!\theta}\Big)f_n^c = 0
\label{eq:NHEK207}
\end{multline}
With $\eta_0$ and $u_0$ given this equation can be solved.

We consider first the simplest case, $u_0=\rm const.$ (and therefore $\eta_0=\rm const.$ as well). Then the general solution for the velocity fluctuation is given by
\begin{multline}
\de u = \sum_{n=0}^\infty \cos{\big(n(\phi+\frac12\,\ln{(1+r^2)})\big)} \\
\big(f^{cc}_n(t-\arctan{r}) \cosh{\Theta_n}+f^{cs}_n(t-\arctan{r})\sinh{\Theta_n}\big)\\ 
+ \sum_{n=1}^\infty \sin{\big(n(\phi+\frac12\,\ln{(1+r^2)})\big)}\\
\big(f^{sc}_n(t-\arctan{r}) \cosh{\Theta_n}+f^{ss}_n(t-\arctan{r})\sinh{\Theta_n}\big)
\label{eq:NHEK208}
\end{multline}
with
\eq{
\Theta_n:=\frac{n}{2}\,\big(\cos\theta+\ln\tan^2\!\frac{\theta}{2}\big)
}{eq:NHEK209}
where the functions $f^{cc}_n$, $f^{cs}_n$, $f^{sc}_n$ and $f^{ss}_n$ are arbitrary. At the poles $\theta=0,\pi$ the Fourier modes diverge because of the term $\ln\tan^2\!\frac{\theta}{2}$ appearing in \eqref{eq:NHEK209}. We should discard fluctuations that diverge at some points, because clearly the linearized Ansatz is no longer valid for such contributions. The only admissible fluctuation therefore comes from the zero mode, $n=0$, and the regular part of the fluctuation $\de u$ becomes independent of the polar and azimuthal angles,
\eq{
\de u^{\rm reg} = f_0 (t-\arctan{r})
}{eq:NHEK210}
The arbitrary function $f_0$ is not fixed by any requirement so far. It is remarkable that there can be no oscillations in the angular coordinates. If we require that the fluctuations vanish at some initial time $t=t_0$ then $f_0(t_0-\arctan{r})$ must vanish for all values of $r$, which implies that $f_0$ must vanish in the interval $(t_0-\pi/2,t_0+\pi/2)$. The discussion above points towards a rigidity statement: it seems unlikely that there are any non-trivial linearized perturbations for a background velocity profile with $u_0=\rm const.$

We consider now generic functions $u_0$, $\eta_0$. In that case it is quite difficult to solve the second order differential equation \eqref{eq:NHEK207}. Moreover, it may well be that again all Fourier modes have to vanish because of regularity requirements. We focus therefore on the zero mode $n=0$. 
\eq{
\de u = A(\theta) f_0(t-\arctan{r})
}{eq:NHEK211}
Then \eqref{eq:NHEK207} simplifies to
\eq{
A'' + A' \Big(2\frac{u_0'}{u_0}+\frac{\eta_0'}{\eta_0}+\frac{(3-\cos^2\!\theta)\cot\theta}{1+\cos^2\!\theta}\Big)=0
}{eq:NHEK107}
where prime denotes differentiation with respect to $\theta$. One solution is given by $A=\rm const.$ The other solution is obtained by defining $y:=u^2_0\eta_0 A'$, which leads to the first order differential equation
\eq{
y'+y\,\frac{(3-\cos^2\!\theta)\cot\theta}{1+\cos^2\!\theta}=0
}{eq:NHEK108}
The general solution for the function $A$ is given by
\eq{
A(\theta)=A_0 + A_1 \int^\theta \frac{1+\cos^2\!x}{u_0^2(x)\eta_0(x)\sin{x}}\,dx
}{eq:NHEK109}
If either the velocity profile or the viscosity has a zero for some value of $\theta$ the second term in \eqref{eq:NHEK109} has a singularity and for consistency we must set $A_1=0$. Even if $u_0$ and $\eta_0$ are non-vanishing and regular throughout the whole spacetime we may have to set $A_1=0$ for consistency. An example is the simple solution $\eta_0=\rm const.$, $u_0=\rm const.$, which leads to an integral in the second term of \eqref{eq:NHEK109} that diverges at the poles $\theta=0,\pi$. Typically, the only allowed fluctuations have $A_0\neq 0$, $A_1=0$. Without loss of generality we may set $A_0=1$. Consistent linearized perturbations for the zero mode $n=0$ are therefore given by
\eq{
\de u^a = u^a\,f_0(t-\arctan{r})
}{eq:fluctuations}
By the same token as before we obtain a rigidity result: if $\de u^a$ is supposed to vanish at an initial time $t=t_0$ the function $f_0$ must vanish in the open interval $(t_0-\pi/2,t_0+\pi/2)$. 

From the discussion above we may conjecture that there are no physically relevant first order scaling fluctuations. If this is true then the first non-trivial fluctuations can emerge only at second order.

\section{Astrophysics and Kerr/CFT}
\label{sec:6}

In this section we summarize and interpret our results, and put them into the perspective of the diverse literature. 

In section \ref{sec:1} we considered the black hole GRS1915+105, whose spin is nearly maximal. The dynamics close to the innermost stable circular orbit (ISCO) could therefore be well described by considering viscous fluid solutions on an extremal Kerr background in the near horizon region. A non-standard feature of the fluid is that on the ISCO its velocity vector should not be timelike but lightlike, since the ISCO coincides with the extremal horizon. In section \ref{sec:2} we introduced the near horizon extremal Kerr (NHEK) geometry \eqref{eq:NHEK7}. The NHEK spacetime is obtained by considering a double limit of the Kerr spacetime: near horizon approximation and near extremality of the Kerr black hole. We reviewed geodesics of timelike and lightlike test particles in the NHEK background for fixed polar angle $\theta=\pi/2$. We generalized these results to arbitrary polar angles $\theta$ and polar velocities $u^\theta$ and identified a third constant of motion $\mathcal D$ \eqref{eq:geotheta1}, analogous to the Kerr case \cite{Carter:1968rr}. In section \ref{sec:3} we constructed exact solutions for a timelike perfect fluid \eqref{eq:pfsol} on the NHEK background, but we had difficulties in including viscosity. Since the NHEK geometry describes the physics on the ISCO, physically relevant results can be obtained if we demand that the velocity vector be lightlike instead of timelike. Consequently, our next step consisted in finding exact solutions for a viscous null fluid in section \ref{sec:4}. We started from the conservation equation of the energy momentum tensor for a viscous null fluid, presented in \eqref{eq:NHEK47}. The results that we obtained for the velocity profile, density, pressure and viscosity are presented in \eqref{eq:vfsol}. For any given velocity distribution we predict uniquely the viscosity function, up to rescaling \eqref{eq:NHEK43}. We added an electromagnetic field \eqref{eq:NHEK36}. Consequently, the energy momentum tensor receives an additional contribution given by \eqref{eq:NHEK40}. For the velocity profile \eqref{eq:vfsol}, one particular solution for the gauge field that does not induce any current takes the form \eqref{eq:NHEK55}. In section \ref{sec:5} we considered first order perturbations around the background solution \eqref{eq:vfsol}. Our assumptions \eqref{eq:NHEK111} -- \eqref{eq:NHEK83} led us to separate the perturbations into scaling and vertical fluctuations \eqref{eq:NHEK84}. We obtained some rigidity results from which we concluded that there are no physically significant first order perturbations. 

In summary, it is generally not possible to find exact solutions to the viscous fluid equations in the presence of electromagnetic fields. We were able to succeed due to the simplifications implied by the highly symmetric NHEK background \eqref{eq:NHEK7}.

We address now potential applications in astrophysics. A quantity of interest in numerical simulations of accretion disks is the viscous torque (cf.~e.g.~\cite{accretion,accretion2}; electromagnetic torque \cite{Ruffini:1975ne} does not arise for our solution). Namely, some boundary condition should be imposed on the viscous torque, and it is debatable what is the appropriate boundary condition at the ISCO. It may seem natural to demand vanishing viscous torque at the ISCO, see e.g.~\cite{Abramowicz:2008bk} for a review, but this is not necessarily a good condition (see for instance \cite{Anderson:1988gd}, where it is also shown that viscous torque can change its sign). Viscous torque in our case is determined by one component of the viscosity tensor \eqref{eq:vistens}, namely by $\Pi^r_\phi$, integrated over a surface of constant radius. This result can be derived by full analogy to the Kerr case \cite{Anderson:1988gd}. The integrated conservation equation for the angular momentum flux $J_1$ consists only of one term, namely the viscous torque integrated along the ISCO boundary, see \eqref{eq:NHEK408}. Our result \eqref{eq:NHEK409} shows that for the NHEK background the viscous torque cannot change its sign. The result \eqref{eq:NHEK409} also provides the appropriate boundary condition for the viscous torque as a function of the velocity profile $u_0(\theta)$. It could be interesting to implement \eqref{eq:NHEK409} as boundary condition for the viscous torque in numerical simulations of viscous fluids on (nearly) extremal Kerr backgrounds. 

It is not necessarily straightforward to translate physical results obtained in the NHEK geometry to corresponding results in the Kerr geometry. In this context it is worthwhile mentioning that the NHEK coordinates \eqref{eq:NHEK7} were obtained from the Boyer-Lindquist coordinates \eqref{eq:kerr} by virtue of two coordinate transformations, none of which involved the polar angle $\theta$. Thus, the $\theta$-dependence of physical quantities like velocity, viscosity or electromagnetic fields is susceptible to a direct translation into the Kerr geometry. This is not true for radial dependences. Since the whole NHEK geometry corresponds to the ISCO in the original Kerr geometry it is not even clear what a radial dependence in NHEK coordinates means in terms of the original geometry. Therefore, we should trust only statements that involve polar angles, like the result for the viscous torque discussed in the previous paragraph or the relation \eqref{eq:NHEK43} between velocity and viscosity.

In our discussion we have encountered some special values for the polar angle $\theta$: the values $\theta\approx 0.82$ (and  $\theta\approx 2.32$) separate the ``ergo-region'' where the Killing vector $\partial_t$ becomes spacelike from the ``normal region'', see Fig.~\ref{fig:1}. The current related to the energy flux $J_0$ \eqref{eq:NHEK402} has the same asymptotic behavior as the Killing vector $\partial_t$, and it is lightlike for the two critical angles mentioned above. The value $\theta\approx 0.955$ leads to vanishing Chern--Pontryagin density \eqref{eq:NHEK13} and vanishing geodesic deviation for a timelike perfect fluid \eqref{eq:geodev2}. It could be interesting to look for experimental signatures close to these special polar angles in real data. 

An interesting effect in the rich physics of accretion disks are quasi-periodic oscillations (QPOs). As mentioned in section \ref{sec:5}, we have not found any QPOs for perturbations around our viscous null fluid solution. This is consistent with the data selection procedure that led to the spin results for the black hole GRS1915+105 \cite{McClintock:2006xd}: the first of the three criteria employed by McClintock et al.~states explicitly that QPOs must be absent (or very weak). 

Of course, all our results are based upon certain assumptions. We discuss now briefly which of them could be relaxed and how this would influence our analysis. Our viscous null fluid solution \eqref{eq:vfsol} describes only the physics at the ISCO of an extremal Kerr black hole. If we drop both assumptions we are back to viscous fluids on a general Kerr background, which is too complicated for analytic studies. It may be possible, however, to use perturbation theory, taking our viscous fluid solution \eqref{eq:vfsol} as the leading order result. Perhaps the ``near NHEK''-geometry constructed recently \cite{Bredberg:2009pv} can provide the suitable next-to-leading order background geometry. Another important assumption was that the velocity vector is lightlike. We motivated this physically by arguing that particles at the ISCO of an extremal Kerr black hole can only move with the speed of light, but slightly outside the ISCO the fluid should have a timelike velocity vector. Thus, for certain perturbative considerations the timelike fluid may provide a better starting point. However, we have not succeeded in finding exact viscous fluid solutions, unless the velocity is lightlike. The derivation of our exact solution \eqref{eq:vfsol} required some additional assumptions, all of which were motivated. Some of them could be relaxed, and therefore we collect here our assumptions for future generalizations: we assumed vanishing angular momentum; we assumed that the polar component of the velocity vector is small as compared to the other components, and hence set $u^\theta=0$; we assumed that all scalar quantities depend only on the polar angle $\theta$ \eqref{eq:NHEK34} so that a matching to scalar quantities in the Kerr geometry is possible. Another interesting generalization would be the construction of solutions with a globally causal current. We have found only solutions with electromagnetic fields where the current is not causal globally or where the current vanishes. Finally, our results on first order perturbations and our conclusion that QPOs are absent are based on the assumptions \eqref{eq:NHEK111} and \eqref{eq:NHEK83}. Dropping either or both of these assumptions might lead to different results.

Finally, we address briefly some (rather superficial, but expandable) relations to the conjectured (extremal) Kerr/CFT correspondence by Guica et al.~\cite{Guica:2008mu}. If the conjecture holds then the near-extreme black hole GRS 1915+105 is approximately dual to (the chiral half of) a conformal field theory (CFT), with left central charge $c_L \sim 2 \times 10^{79}$. The Kerr/CFT correspondence starts from the same assumptions as the present work: a double limit is taken that leads to the NHEK geometry \eqref{eq:NHEK7}. The chirality of the dual CFT implies that all physical excitations must be lightlike, which concurs with our assumption that the velocity profile is lightlike. It may be worthwhile to recover some of the results above from the CFT perspective, like the prediction of the viscosity function from the velocity distribution, the specific boundary condition for the viscous torque, the apparent absence of QPOs or the appearance of some critical angles $\theta$.

\acknowledgments

We thank Bruno Coppi, Florian Preis, Paola Rebusco, Paul Romatschke, Dominik Schwarz and Bob Wagoner for discussions.

DG was supported by the project MC-OIF 021421 of the European Commission under the Sixth EU Framework Programme for Research and Technological Development (FP6). During the final stage DG was supported by the START project Y435-N16 of the Austrian Science Foundation (FWF). 

AP was supported by the project AO/1-5582/07/NL/CB, Ariadna ID 07/1301 of the European Space Agency. 

\begin{appendix}

\section{Perfect fluid equations in NHEK}\label{app:A}

Here we collect some useful formulas for timelike perfect fluids, $u^a u_a=-1$, in the NHEK geometry. 
%
We assume  throughout the paper that the velocity 4-vector components depend only on $r$ and $\theta$. With this assumption the equations of motion simplify considerably. The continuity equation \eqref{eq:NHEK22} reads explicitly
\begin{multline}
u^r \partial_r \rho + u^\theta \partial_{\theta} \rho + (P + \rho) \\
\Big(u^\theta \frac{(3\cos^2\!\theta-1)\,\cot\theta}{1+\cos^2\theta} +\partial_r u^r +  \partial_{\theta} u^\theta \Big)=0
\label{eq:app1}
\end{multline}
The momentum equations \eqref{eq:NHEK23} read explicitly
\begin{widetext}
\begin{multline}
u^t (u^r \partial_r P + u^\theta \partial_{\theta} P) + (P + \rho)  \Big(r u^t u^r \frac{4(\cos^4\!\theta+4\cos^2\!\theta-1)}{(1+r^2) (1+\cos^2\theta)} -  u^t  u^\theta \frac{2\sin\theta\cos\theta}{1+\cos^2 \theta}  
-u^r u^\phi \frac{4
\sin^2\theta}{(1+r^2) (1+\cos^2 \theta)^2}  \\
+ u^r \partial_r u^t + u^\theta\partial_{\theta} u^t\Big)=0
\label{eq:momtt}
\end{multline}
\begin{multline}
u^r (u^r\partial_r P + u^\theta \partial_{\theta} P) + \frac{1+r^2}{2M^2(1+\cos^2\theta)}\, \partial_r P + (P + \rho)   \Big ( (u^t)^2\, r(1+r^2) \frac{\cos^4\!\theta+6\cos^2\!\theta-3}{(1+\cos^2\theta)^2} \\  -u^t u^\phi (1+r^2) \frac{4
\sin^2\theta}{(1+\cos^2\theta)^2}  -(u^r)^2 \frac{r}{1+r^2} -
u^r u^\theta  \frac{2\sin\theta \cos\theta}{1+\cos^2\theta}
 + u^r\partial_r u^r + u^\theta \partial_{\theta} u^r\Big)=0
\label{eq:momtr}
\end{multline}
\begin{multline}
u^\theta (u^r\partial_r P + u^\theta\partial_\theta P ) + \frac{1}{M^2 (1+\cos^2\theta)}\, \partial_{\theta} P  + (P + \rho) \Big (-(u^t)^2 \sin\theta \,\big(\frac{1}{1+\cos^2 \theta}  + r^2 \frac{\cos^4\!\theta+2\cos^2\!\theta+9}{2(1+\cos^2\theta)^3}\big) \\
+ \big((u^r)^2 \frac{1}{1+r^2}  - (u^\theta)^2\big) \frac{\sin\theta
\cos\theta}{(1+\cos^2\theta)} - u^\phi \big(16 r u^t  +
u^\phi\big) \frac{\sin\theta \cos\theta}{(1+\cos^2\theta)^3} + u^r \partial_r u^\theta  + u^\theta \partial_{\theta} u^\theta\Big)=0
\label{eq:momttheta}
\end{multline}
\begin{multline}
u^\phi (u^r \partial_r P + u^\theta \partial_{\theta} P) + (P + \rho) \Big(u^t u^r
\big(\frac{1}{1+r^2} - r^2 \frac{\cos^4\!\theta+6\cos^2\!\theta-3}{(1+r^2)
(1+\cos^2\theta)^2}\big) -  2r u^t u^\theta \frac{(\cos^4\!\theta-2\cos^2\!\theta-3)\cot\theta}{1+\cos^2\theta} \\
+  4 u^r u^\phi \frac{r}{(1+r^2) (1+\cos^2\theta)} +   2
u^\theta u^\phi \frac{\cot\theta}{1+\cos^2\theta}  + u^r \partial_r u^\phi  + u^\theta \partial_{\theta} u^\phi\Big)=0
\label{eq:momtphi}
\end{multline}
\end{widetext}
Only four of the five equations above are independent.

\section{Viscous null fluid equations in NHEK}\label{app:B}

The shear viscosity tensor \eqref{eq:NHEK44} for the velocity distribution \eqref{eq:NHEK33} takes the following form:
\begin{subequations}
\label{eq:vistens}
\begin{align}
\Pi_{t\theta} &= M^2\,\eta\,(1+\cos^2\!\theta)\,(1+r^2)\,\partial_\theta u^t\\
\Pi_{t\phi} &= 2M^2\,\eta\,\frac{\sin^2\!\theta}{1+\cos^2\!\theta}\,\big(r(1+r^2)\,\partial_r u^t-2\,u^t\big)\\
\Pi_{rr} &= -\frac12 M^2\,\eta\,\frac{1+\cos^2\!\theta}{1+r^2}\,\big(3\,(1+r^2)\,\partial_r u^t + 2r\,u^t\big) \nonumber \\
& -\frac32 M^4\, \eta\,(1+\cos^2\!\theta)^2\,(u^t)^2\,\partial_r\big((1+r^2)\,u^t\big) \\
\Pi_{r\theta} &= -M^2\,\eta\,(1+\cos^2\!\theta)\,\partial_\theta u^t\\
\Pi_{r\phi} &= 4M^2\,\eta\,\frac{\sin^2\!\theta}{1+\cos^2\!\theta}\,u^t \label{eq:appeq} \\
\Pi_{\theta\theta} &= \frac12 M^2\,\eta\,(1+\cos^2\!\theta)\,\partial_r\big((1+r^2)\,u^t)\\
\Pi_{\theta\phi} &= 0 \\
\Pi_{\phi\phi} &= 2M^2\,\eta\,\frac{\sin^2\!\theta}{1+\cos^2\!\theta}\,\partial_r\big((1+r^2)\,u^t\big) 
\end{align}
\end{subequations}
The remaining half of the components follows from the symmetry, tracelessness and projection properties \eqref{eq:NHEK45}. With the results \eqref{eq:vistens} the continuity equation \eqref{eq:NHEK49} simplifies to
\eq{
-\frac{1}{2} \eta \,(2 r u^t + (1+r^2)\, \partial_r u^t)=0
}{eq:app2}
By integrating \eqref{eq:app2} we obtain that ${\it u^t}$ has the form
\eq{u^t=\frac{f(\theta)}{1+r^2}
} {eq:ut}
For convenience, we define $f(\theta)=\frac{u_0 (\theta)}{1+\cos^2\theta}$; consequently, \eqref{eq:ut} takes the form
\eq{u^t=\frac{u_0(\theta)}{(1+r^2) (1+\cos^2\theta)}
}{eq:ut2}
The ${\it t}$-component of the momentum equations will then reduce to \eqref{eq:NHEK43}. The $\theta$ momentum equation requires the density to be constant, $\rho=\rho_0$ \eqref{eq:vfsol}. The ${\it r}$ and $\phi$ momentum equations are redundant.


\end{appendix}


\end{document}